# Sustainability of ICT hardware procurement in Switzerland

A status-quo analysis of the public procurement sector

Tobias Welz
Institute of Computer Science
University of Bern
Bern, Switzerland
tobias.welz@inf.unibe.ch

Matthias Stuermer
Institute of Computer Science
University of Bern
Bern, Switzerland
matthias.stuermer@inf.unibe.ch

## ABSTRACT

Sustainable procurement requires organizations to align their purchasing behavior with regard to broader goals linked to resource efficiency, climate change, social responsibility and other sustainability criteria. The level of sustainability in Information and Communication Technology (ICT) hardware procurement is analyzed for two reasons: First, ICT hardware belongs to the six key product groups in sustainable procurement. Second, ICT in general is expected as an important enabler for low-carbon economies, providing solutions to reduce Green-House Gas (GHG) emissions. While the opportunities of sustainable procurement are obvious, certain barriers hinder the adoption in day-to-day procurement. With this case study on ICT hardware, three important barriers "lack of clear definitions per product group", "missing market intelligence about sustainable products" and "inflexible procedures and attitudes as barrier for innovative approaches" are discussed. This publication presents an in-depth analysis for sustainable procurement of ICT hardware by the public sector in Switzerland. For this, tender data published on the national procurement platform simap.ch is examined towards sustainability criteria using the Common Procurement Vocabulary (CPV) nomenclature to identify relevant ICT projects. The results indicate the current level of such criteria included in public tenders as well as their determinants. Using two different approaches, only a small number of tenders containing sustainability criteria ranging from most basic to most comprehensive were found. To analyze the overall performance of Swiss public procurement the identified sustainability criteria are benchmarked with available criteria by international key actors in sustainable procurement. Thus, this analysis provides novel insights on how public agencies today take sustainability into account when procuring ICT hardware.

## 1 Introduction

Discussing the global economic importance of sustainable public procurement, 9.5 trillion USD are spent globally per year as stated by the World Trade Organisation (WTO) to promote sustainability [1]. In the European Union (EU) annual government expenditures on works, goods and services account for roughly 2.0 trillion USD, representing 14% of EU GDP [2]. In Switzerland 5.7 billion USD were spent only by the central federal administration in 2018 on works, goods and services [3]. These expenditures reflect the purchasing power of public procurement agencies and their obligation for important contributions towards local, regional, national and international sustainability goals.

In the context that Information and Communication Technology (ICT) equipment is ubiquitously used and the pictured global demand and corresponding environmental and social impacts are that high, as shown later in this section, it becomes important to consider life-cycle impacts in case of procurement [4]. In fact, ICT hardware procurement turns out to be one of the six most important product categories in private and public sustainable procurement [5] and as observed by United Nation (UN) Environment, sustainable ICT hardware procurement is indicated number one priority by national governments globally [6]. This means, that ICT hardware appears most prior in first order categories which are often purchased centrally or for basic day-to-day operations.

Reflecting the understanding of sustainability criteria in ICT hardware procurement, means that only considering energy consumption and corresponding Green-House Gas (GHG) emissions in the use-phase is a "most basic" approach. Moreover, facing global challenges in sustainable procurement demands a "most comprehensive" approach, as for ecological criteria focusing on GHG emissions throughout the life-span, abiotic mineral depletion in manufacture- and use-phase, economic life-span optimization of ICT equipment using concepts of total cost of ownership (TCO) or life-cycle costing (LCC) as well as addressing social criteria on child labor and workers' safety [7].

In order to discuss the effects of ICT in particular, we first look at the general context of the climate debate. The term sustainability is here equated with the goals of the Paris Climate Treaty of 2015 (COP-21) [8]. The Intergovernmental Panel on Climate Change (IPCC) [9] has published a special report on what needs to happen to achieve its goals. For example, the IPCC is calling for a switch to 100% renewable energy in the areas of buildings, transport and power supply, to comply with global $CO_2$ reduction targets by 2050. The aim is to ensure that global warming is limited to 1.5°C. 2015, global ICT energy consumption [10] turned out at 2300TWh. In 2018 consumption level has risen up to 3000TWh, which corresponds to 2000Mt $CO_2$ or 4% of global GHG emissions, the same amount as civil aviation. If this development continues, ICT emissions demand 8% of global GHG emissions in 2025, the same amount as global car emissions. So ICT would become Top 4 environmental impact.

Today, energy consumption of ICT is composed of 45% producing new equipment and 55% using existing equipment [10]. Taking a closer look on the use-phase, end-consumer devices account for 20% (400Mt $CO_2$) and data traffic demand 35% (700Mt $CO_2$) [data centers 19%, network transmission 16%] of global ICT energy consumption in 2017. The growing demand of new ICT equipment [11] is about 10% cumulated annual growth rate (CAGR) between 2017-2022. To draw the picture, Smartphone demand will grow 9% CAGR with over 6 billion devices in 2022, as also 1.2 billion Personal Computer (PC) and 0.79 billion Tablet Computer. In general, networked ICT devices will almost double between 2017-2022, with the number of machine-to-machine interfaces (M2M) rising sharply and PCs becoming slightly less important.

On average, Swiss consumers cause 14t $CO_2$ per year (2015), 7t $CO_2$ are emitted in Switzerland, the remaining part is released abroad, mainly by foreign production [12]. So, total Swiss GHG emissions account for 95Mt $CO_2$ while the domestic emissions account for 48Mt. To comply with the global climate treaty, domestic $CO_2$ emissions of 54Mt (1990) must be mitigated by -40% till 2030. An ambitious scenario by swisscleantech [13] operationalizes necessary GHG reductions to: buildings -68%, transport -68%, industry -48%, agriculture -24% and waste disposal -6%. Mitigation of foreign emissions is demanded -20% by 2030.

As stated by the UN International Resource Panel, ICT is a key-technology for decoupling natural resource use and environmental impacts from economic growth [14]. So far, worldwide material consumption reached 92.1 billion tons in 2017, up from 87 billion in 2015 and a 254 per cent increase from 27 billion tons in 1970 [15].

Illustrating current status of GHG emissions caused by ICT in Switzerland, a study by Hilty and Bieser [16] states domestic $CO_2$ emissions at 2.55Mt in 2015 where 0.84Mt originate from infrastructure and 1.71Mt from end-user devices. This represents 5.3% of domestic GHG emissions and is above the global average of 4% at present.

Further, this study points out mitigation opportunities using ICT, namely decoupling, towards a low-carbon economy by 2025. Doing so, domestic emissions must be reduced by -10.6Mt $CO_2$. Three scenarios for the development of $CO_2$ emissions are elaborated. A pessimistic scenario assumes an increase of domestic emissions through ICT up to 2.83Mt (+8%) and annual decoupling potentials through ICT by 0.72Mt. An expected scenario assumes 2.40Mt (-6%) with a decoupling potential of 2.79Mt. An optimistic scenario claims 2.08Mt (-17%) with annual decoupling potential of 6.99Mt in other industries. Since the level of domestic emissions has to decline to 37.6Mt by 2025, ICT could so far only contribute to achieve 2%, 16% or at maximum 49% of required GHG emission mitigation for Switzerland to achieve global climate treaty goals.

As shown in the Swiss scenario, ICT can initiate decoupling potentials towards a low-carbon economy, but as global ICT energy demand expands rapidly, accounted gains should not turn into an overall backfire. Backfire means negative rebound effect through overcompensation of realized GHG savings in the use-phase. Hint, the manufacture-phase of ICT equipment already caused essential environmental and social burden.

In case of public ICT procurement, the obligation for important contributions towards more sustainability is twofold. First, sustainable ICT hardware infrastructure must be procured, second using ICT for decoupling in other sectors according to their mitigation targets like for buildings, transport and agriculture. For this study the focus lies in the acquisition of sustainable infrastructure. Decoupling opportunities are mentioned were possible e.g. using life-cycle costing.

Especially in Switzerland, where ICT emissions are above global average it becomes obvious that the public administration establishes comprehensive sustainability approaches regarding ICT and at the same time recognize their role as an important lever towards national sustainability goals.

## 2 Literature

The framework for sustainable procurement builds upon several international agreements. Most general the UN 2030 Agenda for Sustainable Development is mentioned with its 17 sustainable development goals (SDG) [17]. Especially SDG 12 - Ensure sustainable consumption and production patterns as well as SDG 8 - Promote sustained, inclusive and sustainable economic growth, full and productive employment and decent work for all, deal with the concept of sustainable procurement as stated in

- 12.7 promote public procurement practices that are sustainable, in accordance with national policies and priorities.
- 8.4 Improve progressively, through 2030, global resource efficiency in consumption and production and endeavor to decouple economic growth from environmental degradation, in accordance with the 10-year framework of Programmes on Sustainable Consumption and Production, with developed countries taking the lead.

The need for action is mentioned according to the progress of Goal 12 in 2019 as in [15], as worldwide material consumption reached 92.1 billion tons in 2017, with the rate of extraction accelerating every year since 2000. Without urgent and concerted political action, it is projected that global resource extraction could grow to 190 billion tons by 2060.

More practically UN Environment initiated a 10 Year Framework of Programmes on Sustainable Consumption and Production Patterns (10 YFP) [18] consisting of six programs of which one is dedicated explicitly to sustainable public procurement. As 10 YFP aims to foster national action plans, it is recognized that especially European countries are working to embed sustainable procurement within environmental, social, and innovative policies providing some of the best examples of good sustainable public procurement (SPP) practices as in [6].

- 10 YFP – 10 Year Framework on Sustainable Consumption and Production Patterns

With ISO 20400 a sustainable procurement framework is established, addressing all kinds of organizations, as every organization has environmental, social and economic impacts [19]. This framework gives key advice on how to strategically establish sustainable procurement within an organization.

- ISO 20400:2017 Sustainable Procurement – Guidance

Within the European Union (EU) most countries have approved national sustainable public procurement action plans that follow policy recommendations from the European Commission as in [6]. From 2012, new directives by the EU seek to ensure greater inclusion of common sustainability goals in the procurement process. These goals include environmental protection, social responsibility, innovation, combating climate change, employment, public health and other social and environmental considerations, as foreseen in the 7th Environment Action Programme (EAP) in 2013 [20] and stated in the EU public procurement directives.

- Directive 2014/23/EU the Concessions Directive
- Directive 2014/24/EU the Public Sector Directive
- Directive 2014/25/EU the Utilities Sector Directive

The above postulated EU requirements are reviewed for echoes in scientific literature and by key actors in SPP regarding examined obstacles, needs and trends. Therefore, the work of Sönnichsen and Clement [21], Marrucci, Daddi and Iraldo [22], as well as Cheng, Appolloni, D'Amato and Zhu [23] were examined as well as UN Environment [6], EU Green Public Procurement [2], and ICLEI [5]. As a result, it is unfortunately not so easy for public agencies to embrace their duty of exemplarity and enact change, despite increased flexibility according to national action plans. They often go a long way to include social and environmental criteria in public contracts and if included they only address product functionalities, rather than the practices of the suppliers behind those products. To make a real change, it is crucial to recognize the impact of the entire production cycle, not just product functionality, and look deeper into the supply chain to consider environmental, ethical and fair business practices. Moreover, public procurement agencies must have the ability to verify that contract provisions are respected, an impossible feature via contract clauses or supplier codes of conduct alone. A "check the box" approach is insufficient and it is time to go beyond the status quo and further adopt these innovative solutions regarding a sustainable supply chain as they are quickly becoming the norm in the private sector. Further, monitoring and measuring sustainable procurement activities is critical meeting organizations' policy requirements. At the same time such a monitoring would represent the variety of activities. Doing so, political mandates, a professional procurement team as well as the knowledge of the financial efficiency of sustainable alternatives are recognized as precondition to enable sustainable public procurement.

In Switzerland, the Agenda 2030 is adopted by federal administration applied in different roles taken by central and decentralized federal agencies based on responsible entrepreneurial behavior as employer, investor, and procurer [24]. Doing so, the federal administration sets for itself the duty of exemplarity and is recognized in this role by cantons and municipalities. The renewal of federal public procurement law in 2019, which explicitly requires the integration of sustainability criteria in public procurement, is fundamental to the above demanded policy requirements. With regard to the implementation of sustainability criteria into day-to-day procurement a handbook on sustainable procurement has been published as guide for procurers [7] that examines 19 product groups, nine of those have the highest priority regarding sustainability. ICT hardware is one of these product groups.

So far it can be noted that ICT hardware is one of the most important product groups in national and international public procurement activities concerning sustainable criteria.

## 3 Method

As opportunities of sustainable procurement are clear, certain barriers hinder the adoption in day-to-day procurement like "lack of clear definitions per product group", "missing market intelligence to get to know about sustainable products" as well as "inflexible procedures and attitudes to opening-up to innovative approaches" [5].

With the following analysis the above mentioned barriers are addressed showing the performance for sustainable public procurement of ICT hardware in Switzerland. Basically, the first challenge is to get access to information about public tenders. As this information is mandatory for all public procurement tenders simap.ch, the official Swiss electronic tendering platform, provides this kind of information but only current notifications of the last three months. As the simap.ch service is established since 2009, our research center has elaborated a unique data platform, IntelliProcure, for consulting large-scale public procurement

**ICT hardware CPV groups**

| Number of entries | | Number of entries | | Number of entries | |
|---|---|---|---|---|---|
| **221** | **Personal Computer** | **238** | **Server** | **86** | **Smartphone** |
| **214** | **30213** | **160** | **48820** | **12** | **32236** |
| 57 | 30213000 Personal computers | 69 | 4882000 Servers | 12 | 32236000 Radio telephones |
| 63 | 30213100 Portable computer | 25 | 4882100 Network servers | **33** | **32252** |
| 63 | 30213200 Tablet computer | 40 | 4882200 Computer servers | 21 | 32252000 GSM telephones |
| 29 | 30213300 Desktop computer | 8 | 4882300 File servers | 6 | 32252100 Hands-free mobile telephones |
| 2 | 30213500 Pocket computers | 15 | 4882400 Printer servers | 6 | 32252110 Hands-free telephones (wireless) |
| **7** | **30212** | 3 | 4882500 Web servers | **41** | **64212** |
| 5 | 30212000 Minicomputer hardware | **78** | **30211** | 41 | 64212000 Mobile-telephone services |
| 1 | 30212100 CPU for minicomputers | 11 | 30211000 Mainframe computer | | |
| 1 | 30211400 Computer configurations | 3 | 30211100 Super Computer | 64 | **Fixed line device** |
| | | 13 | 30211400 Computer Configurations | **42** | **32552** |
| | | 25 | 30211200 Mainframe hardware | 34 | 32552100 Telephone sets |
| | | 25 | 30211300 Computer platforms | 8 | 32552110 Cordless telephone sets |
| **89** | **Computer Monitors** | **1** | 30211500 Central processing units | **22** | **64215** |
| **8** | **38652** | | | 22 | 64215000 IP telephone services |
| 1 | 38652000 Cinematographic projectors | **170** | **Printer** | | |
| 2 | 38652100 Projectors | **170** | **30232** | | |
| 5 | 38652120 Video projectors | 23 | 30232000 Peripheral equipment | | |
| **81** | **30231** | 73 | 30232100 Printers and plotters | | |
| 25 | 30231000 Computer screens and consoles | 44 | 30232110 Laser printers | | |
| 29 | 30231300 Display screens | 8 | 30232130 Colour graphics printers | | |
| 18 | 30231310 Flat panel displays | 12 | 30232150 Inkjet-printers | | |
| 9 | 30231320 Touch screen monitors | 10 | 30232700 Central controlling unit | | |

**Figure 1: ICT hardware CPV groups - identification of relevant product group clusters**

notices from 2009, as well as tender documents from 2017 on, enabling market intelligence for Swiss public procurement.

So far, the data frame consists of more than 70’000 tenders, 400’000 documents (1.2 TB), 4’000 public procurement agencies, as well as 15’000 suppliers. Machine-learning techniques like Elasticsearch [25] are used to provide a full-text search for all documents as well as other Open Source Software is used to set up this platform. The procedure accessing this data frame is described in the work of Stuermer, Krancher and Myrach [26].

Within the simap.ch data set all tenders are assigned to appropriate product groups in order to recognize all kinds of goods and services. For this classification the Common Procurement Vocabulary (CPV) in the latest version from 2008 [27] is used. The purpose of CPV is to standardize the terms used by contracting authorities to describe the subject of contracts. In particular, these standard codes make it easier to boost transparency as well as to set up an information system for public procurement. The CPV code is structured as an 8-digit number that represents several categories as shown in the following example:

- 30000000 - Office and computing machinery
  the first 2-digts 30 represents CPV-Division
- 30200000 - Computer equipment and supplies
  the 3rd-digt 2 represents the CPV-Group
- 30210000 - Data-processing machines (hardware)
  the 4th-digt 1 represents the CPV-Class
- 30213000 - Personal computers
  the 5th-digt 3 represents CPV-Category
- 30213100 - Portable computers
  the 6th-digt 1 represents CPV-Sub-category

In a study by Grandia and Kruyen [28] the Belgian data frame was screened for sustainability criteria. Screening the data frame roughly on the level of CPV-Divisions “basic” sustainability criteria were detected.

The design of the present study is aiming at retrieving sustainability criteria that are product-group specific, distinct “basic” and “comprehensive” sustainability approaches, as well as to assess the level of sustainability on sector level. Before identifying ICT hardware tenders all ICT related CPV product groups were detected based on a set of national [7] and international [5] [29] requirements, explicitly Personal Computer (Desktop Computer, Integrated Desktop Computer, Thin Client, Notebook Computer - including Tablet computer), Small-scale Server, Computer Monitors - including projectors, Smartphones, Fixed line devices, as well as Printer for office purposes. As result, Fig.1 shows all ICT hardware specific CPV groups and how often these categories were mentioned in the period from 2009 - 2019.

| ICT hardware product groups | Number of tender (2018-2019) | Public procurement agencies | | | |
|---|---|---|---|---|---|
| | | Central federal agencies | Decentral federal agencies | Cantonal agencies | Municipal agencies |
| **Personal Computer** | | | | | |
| total tender found | 61 | 1 | 1 | 26 | 33 |
| tender including sustainabilty | 5 | 1 | 0 | 1 | 3 |
| **Server** | | | | | |
| total tender found | 43 | 2 | 1 | 32 | 8 |
| tender including sustainabilty | 10 | 0 | 1 | 8 | 1 |
| **Smartphone** | | | | | |
| total tender found | 12 | 0 | 2 | 8 | 2 |
| tender including sustainabilty | 2 | 0 | 0 | 0 | 2 |
| **Total** | **116** | | | | |

**Figure 2: ICT hardware tenders containing sustainability criteria (2018-2019)**

The scope of this study is defined as follows:

- ICT Hardware as Personal Computer (Desktop Computer, Integrated Desktop Computer, Thin Client, Notebook Computer - including Tablet computer), Small-scale Server, Computer Monitors - including projectors, Smartphone, Fixed line device, as well as Printer for office purposes.
- Swiss public procurement tenders as in 2018 - 2019, screening tender notices and tender documents.
- Screening ICT hardware according to CPV groups as defined in Fig.1.

# 4 Results

In this chapter public procurement tender for ICT hardware in 2018 and 2019 are identified from the Swiss data set. According to the scope of this analysis the three most important ICT hardware product groups "Personal Computer", "Server" and "Smartphone" are screened for this purpose, as this product groups account the most for the increase in global ICT equipment and at the same time represent the largest shares of global ICT energy demand, that represent a marker for environmental and social burden. The following analysis reflects the current status of public procurement in Switzerland for various types of procurement agencies.

## 4.1 Identification of ICT hardware tender containing sustainability criteria

Applying the CPV categories that define ICT hardware (see Fig. 1), tenders are screened manually for sustainability criteria. Figure 2 shows the amount of tenders found for the three mentioned product categories. 116 tenders were identified in total, of which 17 tenders contain sustainability criteria, which corresponds to a quota of 15%. Looking at the product groups individually, Personal Computer contain 5 of 61 tenders (8%), Server 10 of 43 (23%) and Smartphone 2 of 12 (17%).

In case of procurement agencies, cantonal authorities used sustainability criteria in 9 of 66 (14%) projects as municipal authorities have done in 6 of 43 (14%). Agencies on central and decentralized federal level are so far quantitatively negligible due to a small number of found tenders.

However, there are fundamental differences in the volume of acquisitions as well as in the human resources that manage public tenders. In the case of central and decentralized agencies, tenders are usually procured with large volumes as well as managed by professional procurement teams. In contrast, cantonal procurement offices tend to procure medium-volume orders, while rather small tenders are issued on the level of municipalities, with fewer

**Detailed overview of ICT tenders with sustainabiltiy criteria - Personal Computer (Examples)**

Tender no. 1| Federal Office of Information Technology, Systems and Telecommunication

| | |
|---|---|
| *Project-ID* | 165517 |
| *Type of agency* | Central federal |
| *Value/Volume* | 315 Million CHF \| 90'000 Portable computer, 10'000 Desktop computer, 50'000 Monitors |
| *categories of sustainabilty* | Award criteria (AC) \| Selection criteria (SC) |
| *level o f Sustainability* | **comprehensive** |
| **sustainabilty criteria in tender notice** | **sustainabilty criteria in tender documents** |
| AC - Umweltmanagement - Weighting (15%) | AC - Umweltmanagement - Weighting (15%) |
| AC - Energieeffizienz - Weighting (15%) | AC - Energieeffizienz - Weighting (15%) |
| AC - Energy Star 6.1 | AC - Energy Star 6.1 |
| AC - Energy Star for Displays 7.0 | AC - Energy Star for Displays 7.0 |
| | SC - Rücknahme von Altgeräten (VREG) |
| | SC - End-of-life Management |
| | Blauer Engel |
| | Lifecycle Costing |

Tender no. 2| University of Zurich

| | |
|---|---|
| *Project-ID* | 175484 |
| *Type of agency* | Cantonal |
| *Value/Volume* | 2.2 Million CHF \| 2000 Workstations |
| *categories of sustainabilty* | none |
| *level o f Sustainability* | **basic** |
| **sustainabilty criteria in tender notice** | **sustainabilty criteria in tender documents** |
| Life Cycle Management | Life cycle management |
| | Energy Star |

Figure 3: Overview of ICT tenders containing sustainability criteria – Personal Computer (2018-2019)

**Detailed overview of ICT tenders with sustainabiltiy criteria - Server (Examples)**

Tender no. 8 | High Court of Canton Zurich

| | |
|---|---|
| *Project-ID* | 186914 |
| *Type of agency* | Cantonal |
| *Value/Volume* | 0.3 Million CHF \| Mainframe hardware |
| *categories of sustainabilty* | Award criteria (AC) \| Selection criteria (SC) |
| *level o f Sustainability* | **comprehensive** |
| **sustainabilty criteria in tender notice** | **sustainabilty criteria in tender documents** |
| AC - Nachhaltigkeit - Weighting (5%) | |
| | SC - Grundsatz der Nichtdiskriminierung |
| | Gleichbehandlung von Mann und Frau |
| | Beiträge zu Nachhaltigkeit |
| | Zertifikate sind beizulegen |

Tender no. 9 | Canton of Geneva

| | |
|---|---|
| *Project-ID* | 190176 |
| *Type of agency* | Cantonal |
| *Value/Volume* | n.a. \| Network server |
| *categories of sustainabilty* | Award criteria (AC) |
| *level o f Sustainability* | **comprehensive** |
| **sustainabilty criteria in tender notice** | **sustainabilty criteria in tender documents** |
| AC - Contribution de l'entreprise au développement durable - Weighting (6%) | not available |

Figure 4: Overview of ICT tenders containing sustainability criteria – Server (2018-2019)

employees and less available know-how regarding sustainability. The product category "Personal Computer" shows that at the communal level it is mainly small and medium-sized tenders for workstations, school ICT and mobile clients that are procured, whereas at federal level only large tenders for workstations were found. The high amount of tenders at the municipal level also reflects the need for consulting services, as invitations for tenders are often formulated unspecific for the items to be procured.

## 4.2 Extraction of sustainability criteria for Personal Computer, Server and Smartphone

In order to identify sustainability criteria, all notifications for the respective tenders are screened, mainly tender notices and corresponding tender awards. In case sustainability criteria were discovered in these categories, additional tender documents were examined.

A distinction is made for three categories of sustainability criteria: Award criteria (AC), Selection criteria (SC) and simple nominations. Award categories indicate to what extent the category influences the award of the contract; the higher the weighting, the stronger the influence. Selection criteria are mandatory criteria that must be fulfilled to evaluate candidates' suitability. Simple nominations identify key words or phrases that can be part of award and selection criteria or simply nominations that help to describe sustainability in more detail.

In order to better differentiate found sustainability criteria, general sustainability criteria and product-group specific sustainability criteria are distinguished. For orientation, several practical guides as in [5] [6] [2] give further advice on the international context, as in [7] reflects the Swiss context.

General sustainability criteria nominations ask for more than common practice or explicitly innovative approaches. As an example proper disposal of goods according to the law is social consensus and therefore not recognized as sustainability criteria. Establishing recycling (closed loop resource use) or second-hand practices (reuse of products) on the other hand, is recognized as sustainability criteria.

In case sustainability criteria are recognized in award and selection criteria, this is considered as a "comprehensive" approach. Whereas

**Detailed overview of ICT tenders with sustainabiltiy criteria - Smartphone**

Tender no. 16 | City of Geneva

| | |
|---|---|
| Project-ID | 176963 |
| Type of agency | Municipal |
| Value/Volume | 0.17 Million CHF \| 1600 Smartphones |
| categories of sustainabilty | Award criteria (AC) |
| level o f Sustainability | **comprehensive** |

| **sustainabilty criteria in tender notice** | **sustainabilty criteria in tender documents** |
|---|---|
| AC - Equité sociale - Weighting (5%) | AC - Equité sociale - Weighting (5%) |
| AC- Performance environnementale - Weighting (5%) | AC- Performance environnementale - Weighting (5%) |
| | Politique environnementale<br>certification qualité<br>le domaine environnemental<br>ISO 14001<br>mesures prises en matière d'économie d'énergie<br>politique d'achat préservant l'environnement<br>mesures prises pour limiter les risques d'atteinte à l'environnement |
| | certification qualité officielle dans le domaine social<br>égalité des chances, personnes en réinsertion, handicap<br>démarche concernant l'information et formation sur le développement durable<br>l'égalité entre femmes et hommes.<br>Eco-Entreprise |

Tender no. 17 | Zurich city police

| | |
|---|---|
| Project-ID | 185640 |
| Type of agency | Municipal |
| Value/Volume | 5.5 Million CHF \| 1850 Smartphones , 1400 Tablet computer, Mobile-phone services |
| categories of sustainabilty | Award criteria (AC) |
| level o f Sustainability | **comprehensive** |

| **sustainabilty criteria in tender notice** | **sustainabilty criteria in tender documents** |
|---|---|
| AC - Rückkauf - Weighting (14%) | AC - Rückkauf - Weighting (14%) |
| AC- Entsorgung - Weighting (4%) | AC- Entsorgung - Weighting (4%) |
| | Life Cycle der Geräte zwischen 4-5 Jahren<br>Reparatur, Austausch, Entsorgung<br>Reparatur der Geräte<br>Rückkauf von funktionstüchtigen Geräten<br>fachgerechte Entsorgung |
| | Zertifikat Umwelt-Management<br>ökologische Nachhaltigkeit |
| | Zertifikat Sozial-Management<br>Abschaffung Kinderarbeit<br>ILO-Konvention 100<br>ILO-Konvention 111<br>Beseitigung Zwangsarbeit<br>Verbot Diskriminierung in Beschäftigung |

Figure 5: Overview of ICT tenders containing sustainability criteria – Smartphone (2018-2019)

sustainability criteria that are only loosely mentioned without influencing the decision, reflect a most "basic" approach.

Figure 3-5 show all sustainability criteria found for each of the three ICT hardware product groups. In sum 17 tenders were identified containing ecological and social criteria, six tenders with a most "basic" approach and eleven with a most "comprehensive" approach. As displayed, sustainability criteria were identified for the French and German-speaking areas of Switzerland. Recognizing sustainability criteria, whether basic or comprehensive, in the general tender notice, leads to further screening of attached tender documents. Thus, if for example a most basic approach is found in the tender notice, the same terms are found in the attached tender documents as well as additional terms that better describe the perceived sustainability approach.

## 4.3 Classification of sustainability criteria

Further, a classification of all recognized sustainability criteria discussing the allocation of product-group specific and general sustainability, leads to a cluster of ecological and social criteria (see Fig.6-7). The terms are ordered ascending from general to specific sustainable criteria. For the ecological criteria terms were even clustered into four different levels –general – technique-specific – ICT-specific – product-group-specific–. Hint: this cluster is only valid for the defined ICT hardware CPV groups. Social criteria are not further clustered. With this classification at hand it is possible to search for ICT hardware tenders for any given timeframe and also to evaluate "basic" or "comprehensive" sustainability approaches.

Further German and French sustainability criteria were merged and also translated into their respective English expressions (see Fig. 8-9). With this very list of sustainability keywords, it is possible to further establish queries that allow to search for the content wise presence of keywords of the respective clusters.

## 4.4 Identification of ICT hardware tender – using list of sustainability keywords and extending CPV range

The now identified sustainability criteria are used in a query screening the data frame to check if more ICT hardware tender are found compared to the handy selection. First, such a query is performed for the CPV-range as defined in the scope. Then the grid is extended on the level of CPV-Class and CPV-Group.

In general, more ICT hardware tenders containing sustainability criteria were found. Within the same CPV-range three additional projects, on CPV-Class level four and on CPV-Group level fourteen further projects were identified.

| | ICT hardware product group - **Personal Computer**: sustainabilty criteria in tender notice | Personal Computer: sustainabilty criteria in tender documents | ICT hardware product group - **Server**: sustainabilty criteria in tender notice | Server: sustainabilty criteria in tender documents | ICT hardware product group - **Smartphone**: sustainabilty criteria in tender notice | Smartphone: sustainabilty criteria in tender documents |
|---|---|---|---|---|---|---|
| | *ecological criteria* | | *ecological criteria* | | *ecological criteria* | |
| **general** | | ökologische Nachhaltigkeit<br>Umweltschutz | Nachhaltigkeit<br>Ecologie<br>Environnement<br>composante environnementale au développement durable<br>Contribution de l'entreprise au développement durable | Nachhaltigkeit | Performance environnementale | ökologische Nachhaltigkeit<br>Performance environnementale<br>Politique environnementale<br>politique d'achat préservant l'environnement<br>mesures prises pour limiter les risques d'atteinte à l'environnement |
| **technique-specific** | Umweltmanagement<br>Energieeffizienz | Umweltmanagement<br>ISO 14001 | consommation énergetique | ressourceneffizienteren Betrieb<br>Energieeffizienz<br>label efficacité de la consommation énergétique | | Zertifikat Umwelt-Management<br>ISO 14001<br>certification qualité le domaine environnemental |
| **ICT-specific** | Life-Cycle Management | Lifecycle<br>Life-Cycle Management<br>Lifecycle Costing<br>Blauer Engel (Label) | | Green IT | Rückkauf | Rückkauf |
| **product-group-specific** | Energy Star 6.1<br>Energy Star for Displays 7.0 | Energy Star (Label)<br>Energy Star 6.1<br>Energy Star for Displays 7.0 | | power usage effectiveness (PUE) | | |

Figure 6: Overview of ecological criteria in ICT hardware product groups (2018-2019)

| | ICT hardware product group - **Personal Computer**: sustainabilty criteria in tender notice | Personal Computer: sustainabilty criteria in tender documents | ICT hardware product group - **Server**: sustainabilty criteria in tender notice | Server: sustainabilty criteria in tender documents | ICT hardware product group - **Smartphone**: sustainabilty criteria in tender notice | Smartphone: sustainabilty criteria in tender documents |
|---|---|---|---|---|---|---|
| | *social criteria* | | *social criteria* | | *social criteria* | |
| **general** | Lohngleichheit<br>Mann und Frau | Ethikkodex | Ethikodex<br>Sozialkodex | Sozial- Ethikkodex (SEK)<br>Nichtdiskriminierung<br>Mann und Frau<br>Lohngleichheit<br>Mann und Frau | Equité sociale | Equité sociale<br>Beseitigung Zwangsarbeit<br>Abschaffung Kinderarbeit<br>Verbot der Diskriminierung in Beschäftigung<br>égalité des chances, personnes en réinsertion, handicap |
| **specific** | | ILO<br>ILO-Deklaration<br>ILO-Kernarbeitsnorm | composante sociale du développement durable | | | l'égalité entre femmes et hommes<br>ILO-Konvention 100<br>ILO-Konvention 111<br>démarche concernant l'information et formation sur le développement durable<br>Sozial-Management<br>certification qualité officielle dans le domaine social<br>Eco-Entreprise (Label) |

Figure 7: Overview of social criteria in ICT hardware product groups (2018-2019)

## 4.5 Comparison of sustainability criteria with sustainability requirements of key actors in SPP

Assessing the level of sustainability, found sustainability criteria are compared to sustainability requirements by key actors in SPP as in [5] [2] [29]. Figure 8-9 show on the left-side criteria retrieved from Swiss ICT hardware tenders and on the right-side requirements retrieved from SPP key actors.

For ecological criteria a vast amount of ICT-specific criteria match, even some keywords were additionally found in the tenders. Regarding the product-group specific criteria some ICT-labels were mentioned in Swiss tenders, but the overall performance is rather low.

Examining social criteria Swiss tender perfectly match on the level of general working conditions but perform rather poor for the specific sustainable criteria on working conditions for raw materials and in manufacture. As no respective labels nor initiatives were found this particular area looks like a blind spot.

| ecological criteria in German tender | ecological criteria in French tender | ecological criteria in English (translated) | | ecological requirements by key actors |
|---|---|---|---|---|
| Nachhaltigkeit<br>ökologische Nachhaltigkeit<br>Umweltschutz | Ecologie<br>Environnement<br>Performance environnementale<br>Politique environnementale<br>politique d'achat préservant l'environnement<br>composante environnementale au développement durable<br>Contribution de l'entreprise au développement durable<br>mesures prises pour limiter les risques d'atteinte à l'environnement | sustainability<br>ecological sustainability<br>ecology<br>environmental protection<br>environmental policy<br>environemtnal purchasing policy<br>environmental component to sustainable development<br>Company's contribution to sustainable development<br>measures to limit environmental damage | general | |
| Umweltmanagement<br>Zertifikat Umwelt-Management<br>ISO 14001<br>ressourceneffizienteren Betrieb<br>Energieeffizienz | certification qualité le domaine environnemental<br>ISO 14001<br>consommation énergetique<br>label efficacité de la consommation énergétique | environmental management<br>certified environmental management<br>ISO 14001<br>resource-efficient operation<br>energy consumption<br>energy efficiency<br>energy efficiency label | technique-specific | end of life management<br>low power mode |
| Lifecycle<br>Life-Cycle Management<br>Rückkauf<br>Green IT<br>Lifecycle Costing<br>Blauer Engel (Label) | | Lifecycle<br>Life-Cycle Management<br>Buy-back<br>Green IT<br>Lifecycle Costing<br>Blue Angel (Label) | ICT-specific | Total Costs of Ownership<br>Lifecycle Costing<br>life time extension<br>EU Ecolabel (Label)<br>EPEAT (Label)<br>Blue Angel (Label) |
| Energy Star (Label)<br>Energy Star 6.1<br>Energy Star for Displays 7.0<br>power usage effectiveness (PUE) | | Energy Star (Label)<br>Energy Star 6.1<br>Energy Star for Displays 7.0<br>power usage effectiveness (PUE) | product-group-specific | TCO Certified, generation 8 (Label)<br>Nordic Ecolabel 7.4 (Label)<br>Energy Star Computers 6.1 (Label)<br>Blue Angel UZ 78a (Label)<br>power usage effectiveness (PUE)<br>energy usage effectiveness (EUE)<br>NSF/ANSI 426-2018 (Label)<br>Energy Star Enterprise servers (Label)<br>Blue Angel UZ 171 (Label)<br>UL 110 Edition 2 (Label)<br>Blue Angel UZ 106 (Label) |

Figure 8: Comparison of found ecological criteria with requirements by key actors in SPP (2018-2019)

| social criteria in German tender | social criteria in French tender | social criteria in English (translated) | | social requirements by key actors |
|---|---|---|---|---|
| Ethikodex<br>Sozialkodex<br>Sozial- Ethikkodex (SEK)<br>Nichtdiskriminierung von Mann und Frau<br>Beseitigung Zwangsarbeit<br>Verbot der Diskriminierung in Beschäftigung<br>Abschaffung Kinderarbeit<br>Lohngleichheit Mann und Frau<br>ILO-Deklaration<br>ILO-Kernarbeitsnorm<br>ILO-Konvention 100<br>ILO-Konvention 111<br>Sozial-Management | Equité sociale<br>égalité des chances, personnes en réinsertion, handicap<br>l'égalité entre femmes et hommes<br>démarche concernant l'information et formation sur le développement durable<br>composante sociale du développement durable<br>certification qualité officielle dans le domaine social<br>Eco-Entreprise (Label) | Code of Ethics<br>Social Code<br>Social equity<br>Social Code of Ethics<br>Non-discrimination of men and women<br>Elimination of forced labor<br>Elimination of child labor<br>Non-discrimination in employment<br>Non-discrimination of disabeled people and in rehabilitation<br>Equal pay for men and women<br>training on sustainable development<br>ILO Declaration<br>ILO Convention<br>ILO Convention 100<br>ILO Convention 111<br>social component of sustainable development<br>social management<br>Eco-Entreprise (Label) | working conditons - general | living wage<br>ILO core standards<br>supplier code of conduct<br>Occupational health and safety management systems (OHSMS)<br>Socially responsible public procurement<br>The LANDMARK project |
| | | | working conditons - raw materials & manufacture | Responsible Business Alliance (RBA)<br>Responsible Labor Initiative (RLI)<br>Fair Labor Association (FLA) (Label)<br>Ethical Trading Iniative (ETI) (Label)<br>SA 8000 (Label)<br>Electronis Watch<br>Conflict minerals<br>Sustainable minerals<br>Responsible Minerals Initiative (RBMI)<br>Better Gold Initiative<br>Responsible Cobalt Initiative |

Figure 9: Comparison of found social criteria with requirements by key actors in SPP (2018-2019)

## 5 Conclusions

The particular study-design used in this work allows retrieving specific ICT hardware tenders for the Swiss data frame for a specified amount of product groups from several types of public procurement authorities. It is thus possible to retrieve tenders with distinguished "basic" and "comprehensive" sustainability approaches as well as sustainability criteria (ecological and social) clustered from general to specific. In addition, for the ecological criteria it is even possible to further cluster these into – general – technique-specific – ICT-specific – product-group-specific criteria. Furthermore, a list of keywords is retrieved that defines sustainability in ICT hardware as found in day-to-day procurement as well as from international key actors that define sustainability requirements. These lists of keywords allow assessing sustainability criteria on the overall level of sustainability in ICT hardware. Using appropriate queries even more sustainability tenders could be found as by hand.

With the obtained knowledge it is possible to use this particular study-design to address other sectors retrieving sustainability criteria in general or assessing the level of sustainability. It would be most effective doing this for constructions, transport and agriculture, as these sectors are crucial towards national climate goals. While retrieving keywords for the construction sector would be rather easy, for transport and agriculture the procedure can be rather complex. Beside the assessment of other sectors, it is in addition possible to examine the progress of sustainability criteria over time using continually updated sustainability requirements. Using the methods of this particular study-design has the limitation that only tender notices and tender awards can be addressed with specific queries. Also, screening tender documents automatically using queries would enable to screen bigger data frames at once.

In the Swiss context, the overall retrieved tenders contain a high level of general sustainability criteria but perform low on specific criteria, in particular requesting international standards. Most ecologically comprehensive approaches were found in the German-speaking part while most comprehensive social approaches were found in the French-speaking part. Neither comprehensive nor basic sustainability approaches were found in the Italian-speaking part.

With regard to the level of know-how or professional procurement teams' federal agencies perform best and municipal agencies perform worst. Closing this pre-procurement performance barrier is key with regard to all sustainability goals. In general, sustainability criteria are represented in less than 20% of all public tenders and the identified sustainability criteria vary from basic to comprehensive approaches. As some agencies are pioneers with regard to sustainable criteria many criteria could be found at some point.

Regarding the two levers of ICT in public procurement (sustainable ICT hardware and ICT for decoupling) the necessary requirements such as defined sustainability criteria as well as approaches like life-cycle costing are at hand, but not widely used in day-to-day procurement.

When it comes to the procurement of sustainable ICT hardware only sustainability criteria with regard to product functionalities are addressed. The second lever of ICT, decoupling or more precisely life-cycle costing, performs very low. This low performance can partly be explained by the need for strategic incorporation in the organizations' sustainability concept which is usually a rather complex undertaking. Overall, the present study-design overcomes important barriers in public procurement: the lack of clarity is addressed through definition of suited CPV-groups, missing market intelligence is addressed through retrieval of sustainable tenders and inflexible approaches are addressed through retrieving best-practice examples as well as sustainability criteria for basic and comprehensive approaches.

Concluding the findings of this study it can be stated that it is possible to assess sustainability performance on sector level, in this case for ICT hardware, as well as comparing national levels of sustainability to international requirements. However, it is key to establish a knowledge transfer throughout all levels of procurement agencies to spread professional procurement know-how in order to benefit from these measures.

## ACKNOWLEDGMENTS

This research is part of the National Research Project NRP 73 Sustainable Economy, Grant Number 407340_172351, and funded by the Swiss National Science Foundation.